\documentclass[12pt]{article}

\usepackage{times}

\usepackage[colorlinks=true,bookmarks=false,citecolor=blue,urlcolor=blue]{hyperref} %pdflatex
\usepackage{amsmath,amssymb, mathrsfs}
\usepackage{graphicx}
\usepackage{float}
\usepackage{xcolor}
\usepackage{soul}

\usepackage{capt-of}

\usepackage[
    backend=biber,
    style=nature,
  ]{biblatex}

\newenvironment{sciabstract}{%
\begin{quote} \bf}
{\end{quote}}

\title{Ultrafast neuromorphic computing with nanophotonic optical parametric oscillators}

\author{Midya Parto$^{1,2,3,*,\dagger}$, Gordon H.Y. Li$^{4,*}$, Ryoto Sekine$^1$, Robert M. Gray$^1$,\\ Luis L. Ledezma$^1$, James Williams$^1$, Arkadev Roy$^1$, Alireza Marandi$^{1,4,\dagger}$\\
\\
\normalsize{$^1$Department of Electrical Engineering, California Institute of Technology, Pasadena, CA 91125, USA.}\\
\normalsize{$^2$Physics and Informatics Laboratories, NTT Research, Inc., Sunnyvale, California 94085, USA.}\\
\normalsize{$^3$CREOL, The College of Optics and Photonics, University of Central
Florida, Orlando, FL, USA.}\\
\normalsize{$^4$Department of Applied Physics, California Institute of Technology, Pasadena, CA 91125, USA.}\\
\\
\normalsize{$^*$These authors contributed equally.}\\
\normalsize{$^\dagger$marandi@caltech.edu}\\
\normalsize{$^\dagger$midya.parto@ucf.edu}
}

\date{}

\begin{document} 


\baselineskip24pt


\maketitle 

\begin{sciabstract}
Over the past decade, artificial intelligence (AI) has led to disruptive advancements in fundamental sciences and everyday technologies. Among various machine learning algorithms, deep neural networks \cite{lecun_deep_2015} have become instrumental in revealing complex patterns in large datasets with key applications in computer vision, natural language processing, and predictive analytics. With the increasing prevalence and adoption of deep learning, the quest for hardware solutions that can efficiently process data in real time with high speeds and low latencies has come to the forefront of research in many fields. On-chip photonic neural networks offer a promising platform that leverage high bandwidths and low propagation losses associated with optical signals to perform analog computations for deep learning \cite{ashtiani_-chip_2022,chen_all-analog_2023,pai_experimentally_2023,wetzstein_inference_2020,shastri_photonics_2021}. However, nanophotonic circuits are yet to achieve the required linear and nonlinear operations simultaneously in an all-optical and ultrafast fashion. Here, we report an ultrafast nanophotonic  neuromorphic processor using an optical parametric oscillator (OPO) fabricated on thin-film lithium niobate (TFLN). The input data is used to modulate the optical pulses synchronously pumping the OPO. The consequent signal pulses generated by the OPO are coupled to one another via the nonlinear delayed dynamics of the OPO, thus forming the internal nodes of a deep recurrent neural network. We use such a nonlinearly coupled OPO network for chaotic time series prediction, nonlinear error correction in a noisy communication channel, as well as noisy waveform classification and achieve accuracies exceeding 93\% at an operating clock rate of $\sim$ 10 $\text{GHz}$. Our OPO network is capable of achieving sub-nanosecond latencies, a timescale comparable to a single clock cycle in state-of-the-art digital electronic processors. By circumventing the need for optical-electronic-optical (OEO) conversions, our ultrafast nanophotonic neural network paves the way for the next generation of compact all-optical neuromorphic processors with ultralow latencies and high energy efficiencies.

\end{sciabstract}

Deep neural networks (DNNs) have revolutionized modern data processing with transformative results across numerous fields ranging from fundamental sciences to automotive and healthcare industries to generative art \cite{vaswani_attention_2017,jumper_highly_2021,trinh_solving_2024}. As deep learning algorithms become more prominent, traditional computing hardware proves to be increasingly less suitable for their implementation. This incompatibility poses a formidable challenge for the future of AI and originates from several characteristics of current computing architectures. The substantial growth of the compute required by recent deep learning models far surpasses the current trend in the advancement of digital electronic processors \cite{sevilla_compute_2022}. The resulting gap becomes even more acute in many practical scenarios that require real-time processing where ultralow latencies are critical. In addition, as algorithms become more complex and datasets grow larger, the energy required to perform computations surges, resulting in serious financial and environmental consequences. Providing viable solutions to these challenges has brought the quest for efficient and high-speed computing hardware to the forefront of research across various disciplines to drive disruptive AI technologies.

Against this backdrop, photonic neural networks (PNNs) have recently emerged as promising candidates that can harness high bandwidths offered by optical systems to provide ultrafast operation while maintaining high energy efficiencies facilitated by the low-loss propagation of light \cite{wetzstein_inference_2020,shastri_photonics_2021,mcmahon_physics_2023}. In particular, recent advances in nanofabrication and photonic integrated circuits (PICs) have provided a fertile ground for implementing light-based AI architectures on compact chip-scale devices \cite{feng_integrated_2024}. DNNs consist of many layers of linear operations such as matrix multiplications or convolutions interleaved by nonlinear activation functions. Existing integrated PNNs utilize various architectures such as Mach-Zehnder inteferometers \cite{shen_deep_2017} as well as microring resonator arrays to accelerate the linear operations necessary for deep learning  \cite{ashtiani_-chip_2022,pai_experimentally_2023,seyedinnavadeh_determining_2024}. Typically, opto-electronic components provide the nonlinear activation functions in PNNs. Alternatively, one can realize the nonlinear operations by using all-optical nonlinearities \cite{chen_deep_2023} such as phase change materials \cite{feldmann_all-optical_2019,feldmann_parallel_2021,bruckerhoff-pluckelmann_event-driven_2023} or optical parametric processes \cite{li_all-optical_2023,wright_deep_2022}. This latter approach can therefore eliminate the need for successive optical-electronic-optical (OEO) conversions and provide a route for true all-optical integrated neural networks with ultrashort latencies beyond those attainable by hybrid optoelectronic architectures. Yet, despite significant progress in PNNs \cite{lvovsky2022hybrid}, achieving a unifying platform that simultaneously provides both linear multiply-accumulate (MAC) operations and ultrafast nonlinear activation functions in nanophotonic circuits has remained elusive.

\begin{figure}
    \centering
    \includegraphics[width=0.9\textwidth]{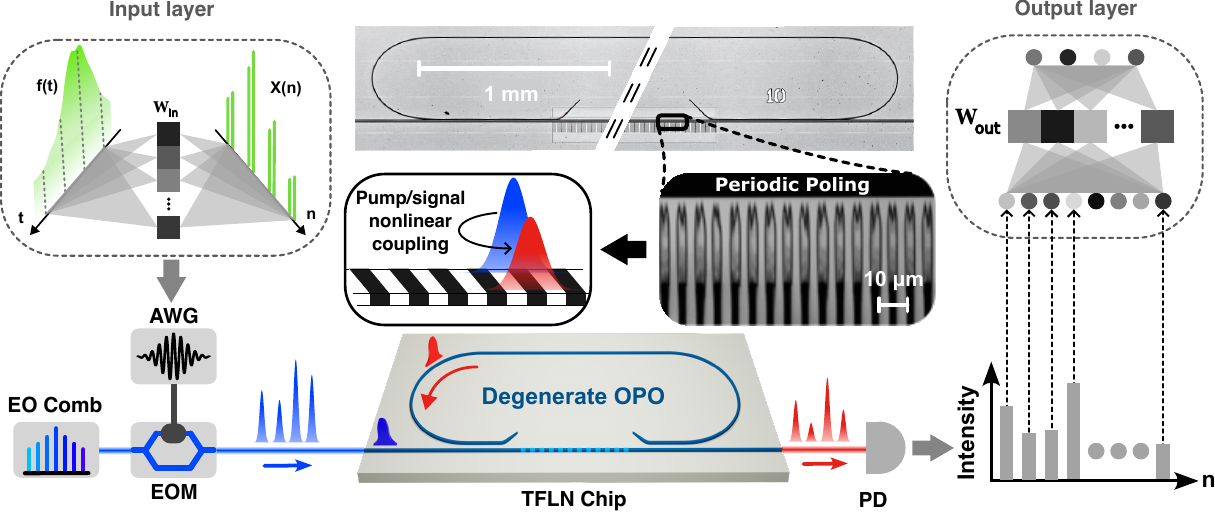}
    \caption{\textbf{Nanophotonic OPO-based neuromorphic processor.} Schematic view of the nanophotonic OPO utilized in our experiements for neuromorphic computing. The top panel depicts the microscope image of the device (more information about the chip design can be found in \cite{ledezma_octave-spanning_2023}). The central part of the racetrack resonator is removed for clarity. The structure is designed in such a way that only the degenerate signal/idler resonate inside the cavity. At the input, the optical pump generated from an EO comb at $f_{\text{rep}} \approx 10 \text{ GHz}$ and a center wavelength of $\sim 1045\text{ nm}$ is modulated by an EOM. The modulating signal is generated by an AWG that applies discrete data $X(n)$ at $\sim 10 \text{ GSa/s}$. This data represents a time series that is obtained by sampling an arbitrary function  $f(t)$ in time, followed by masking with a randomized set of input weights $\textbf{W}_\text{in}$, thus forming the input layer of our OPO neural network. The inset depicts a  two-photon microscope image of the periodically poled region in the OPO cavity necessary for quasi phase matching (QPM). Within this poled region, the generated signal pulses from previous roundtrips nonlinearly couple to the incoming pump pulses, collectively acting as a nonlinear recurrent network that connects the input layer to the output (the inset schematically displays this nonlinear coupling). At the output, the signal pulses at $\sim 2090\text{ nm}$ are separated from the pump using a fiber-based WDM filter and sent to a fast photodetector. The output layer of our OPONN is then set up by forming a weighted average of these detected pulses.}
    \label{fig:OPONet}
\end{figure}

In this work, we address this challenge and demonstrate an integrated photonic neural network that harnesses coherent optical pulse propagation in conjunction with ultrafast parametric nonlinear processes \cite{soljacic2023biasing,soljacic2024probabilistic} to achieve ultralow-latency operations. To this end, we use a nanophotonic optical parametric oscillator (OPO) fabricated on a TFLN chip to implement a deep recurrent neural network which we dub OPONN. In our scheme, as shown in Fig.~\ref{fig:OPONet}, masked sequential data is used to modulate the optical pulses of an electro-optic (EO) frequency comb used to synchronously pump the OPO. In response, the nanophotonic OPO, which operates at degeneracy~\cite{roy2023visible}, generates signal pulses at the half-harmonic of the input pump. The optical feedback provided by the cavity on one hand and the parametric amplification on the other hand act upon the generated signal fields creating dynamics akin to those associated with neurons in a recurrent neural network. We utilize our OPONN for machine learning on three different tasks: (i) ultrafast chaotic time series prediction involving the archetypal Lorenz and Mackey-Glass systems of equations, (ii) compensating nonlinear distortions in a noisy communication channel, and (iii) waveform classification on noisy signals that are randomly chosen from three different classes of sinusoidal, rectangular, and sawtooth functions. In all cases, the OPONN achieved accuracies exceeding 93\% while operating at $\sim10~\mathrm{GHz}$ clock rates. Remarkably, this clock rate is only limited by the electronic sources responsible for generating the input pump pulses, and is not an inherent limit of our OPO-based neural network, which is operable with femtosecond pulses. By realizing the linear MAC operations as well as the nonlinear activation functions in the optical domain, our implementation also eliminates the need for OEO conversions that are typically required in state-of-the-art PNN architectures.

Figure \ref{fig:OPONet} illustrates the OPO-based neurmorphic processor. At the heart of the processor is an on-chip OPO which is fabricated on an x-cut TFLN wafer with a silica buffer layer. It consists of a dispersion-engineered periodically-poled lithium niobate (PPLN) section that provides broadband quasi phase matching (QPM) between the pump at 1045 nm and the generated signal and idler waves. The cavity is formed by two input/output adiabatic couplers which are designed to only allow the signal/idler modes to resonate within the cavity~\cite{roy2023visible}. Throughout our experiments, the OPO is operated in the degenerate regime. We synchronously pump the OPO using transform-limited pulses with a duration of $\sim2~\mathrm{ps}$ generated by an EO comb with a repetition rate of $\sim10~\mathrm{GHz}$ matching the cavity free spectral range (FSR). The input layer of the neural network is  formed by using an electro-optic modulator (EOM) driven by an arbitrary waveform generator (AWG) operating at $\sim10~\mathrm{GSa/s}$ to prepare the masked sequential data fed into the OPO as the pump pulses. During successive cycles defined by the roundtrip time of the OPO cavity, the input pump pulse and the signal pulse traveling in the cavity overlap in the PPLN region. The signal pulse carries information from the previous signals due to the feedback provided by the cavity. In this respect, the resonator serves as an optical memory. Followed by this coherent interference between the signal and pump pulses, the PPLN element acts as a nonlinear neuron that generates a signal pulse which is a nonlinear function of its inputs, i.e. the current pump pulse and the previous signal pulse. In this respect, the nanophotonic OPO provides an all-optical implementation of a deep recurrent neural network. Finally, the weighted average of the signal pulse intensities resolved by a fast detector forms the output layer of the OPONN (Fig. \ref{fig:OPONet}).

\subsection*{Low latency time domain signal processing using OPONN}

\begin{figure}
    \centering
    \includegraphics[width=1\textwidth]{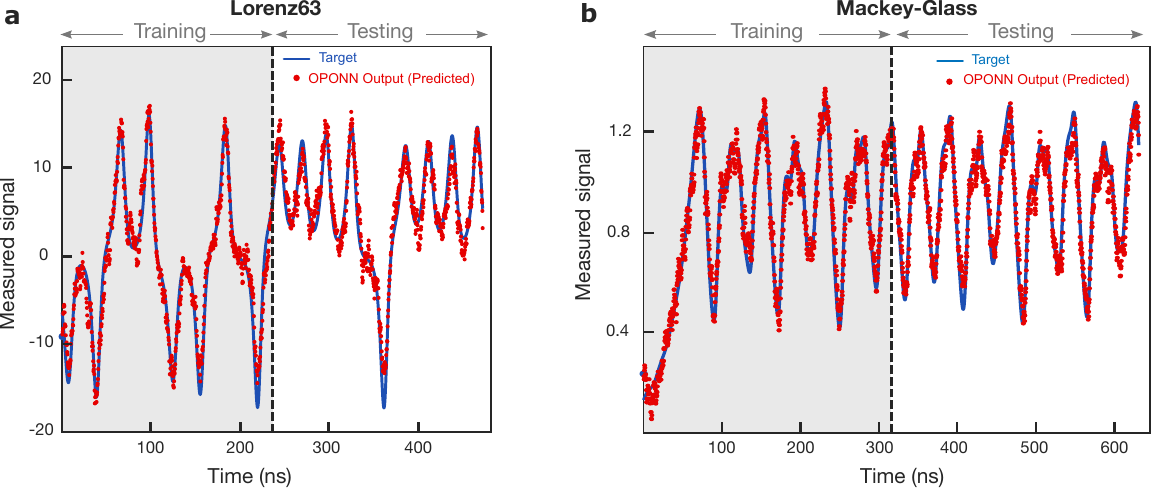}
    \caption{\textbf{Chaotic time series prediction using OPONN. a,} Numerically calculated values of the $x(t)$ signal associated with the Lorenz63 model (target) and the experimentally measured signal at the output of our OPONN (predicted). As shown in the figure, the entire span of the signal with a $\sim 500 \text{ ns}$ duration is split into two equal sections for training (the shaded area) and testing. As evident, the predictions of the OPONN closely matches the target values. \textbf{b,} Similar results for the Mackey-Glass system of equations in the chaotic regime.}
    \label{fig:Chaotic}
\end{figure}

We employed the OPONN to perform machine learning on a variety of benchmark tasks that involve time-domain signals. The first task is to predict the evolution of a chaotic time series generated by the Lorenz63 model proposed by Lorenz in 1963 \cite{lorenz_deterministic_1963} to describe atmospheric convection. This is governed by a three-dimensional system of coupled nonlinear differential equations $\Dot{x}=10(y-x), \Dot{y}=x(28-z)-y, \Dot{z}=xy-8z/3,$ where $x$, $y$ and $z$ denote physical observables associated with the convective current and the temperature variations in different spatial directions. For the set of parameters and the initial conditions chosen here, the Lorenz system exhibits deterministic chaos that is manifested by an aperiodic trajectory in the phase space known as a strange attractor. Under such conditions, the evolution of the observables show a dramatic sensitivity to the initial conditions, hence defying conventional methods of prediction. Our goal here is to train the OPONN to forecast the next time step of the signal $x(t)$ based on its past history. To achieve this, we first sample the input data at a sampling rate of $\sim 10 \text{ GHz}$ to obtain the discrete signal $u(n)$. We then configure the network with $N_\text{in}=3$ and $N_\text{out}=10$ nodes in the input and the output layers, respectively. The input weights defined by the vector $\textbf{W}_\text{in}$ are chosen randomly. We split the entire span of the available data into two equal periods of training and testing. Training is performed in silico by singular value decomposition to obtain the optimal output weight matrix $\textbf{W}_\text{out}$ that results in the least estimation error during the training phase. In contrast to other types of deep neural networks that typically require computationally intensive training techniques such as backpropagation, our neural network can be trained considerably simpler with minimal computation requirements. Figure \ref{fig:Chaotic}\textbf{a} shows measurement results obtained from experiments, where both the target and the predicted signals are displayed. To quantify the performance of the OPONN for this forecasting task, we calculate the normalized mean square error (NMSE) between the target and the predicted time series given as $0.07\pm 0.017$ for this task.

As a second example, we consider another benchmark time series prediction task associated with the Mackey-Glass system (MGS), which was originally developed in the study of physiological control mechanisms that are known to exhibit a host of complex behaviors 
\cite{mackey_oscillation_1977}. The model is described by a class of nonlinear delay differential equations exhibiting dynamics that critically depend on the system parameters. Of particular interest is the scenario when the solutions become chaotic, signaling an abnormal physiological behavior associated with pathological cases. Here, we use our OPO network to forecast the future instances of the resulting chaotic time series. Figure \ref{fig:Chaotic}\textbf{b} depicts the target MGS solution together with predicted values obtained by measuring the OPONN output, indicating a prediction error of $\text{NMSE}=0.06\pm 0.017$.

\begin{figure}
    \centering
    \includegraphics[width=1\textwidth]{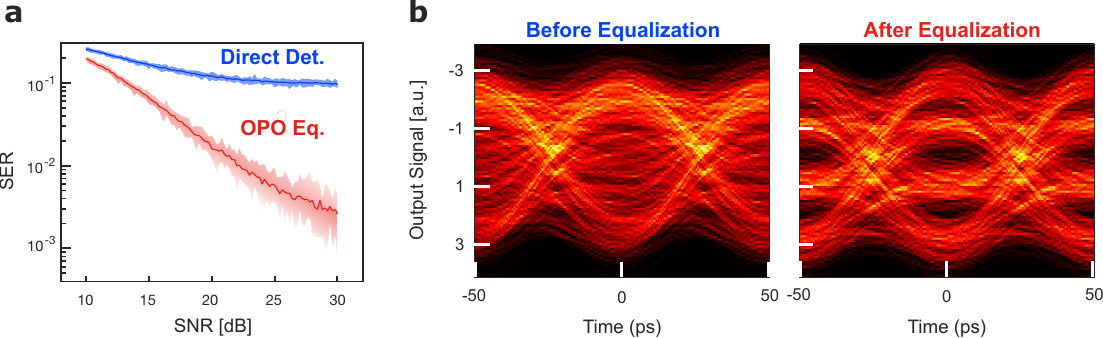}
    \caption{\textbf{Nonlinear channel equalization of PAM4 signals using ultrafast nonlinear response of the OPO network. a,} Simulated SERs obtained using direct detection versus those achieved after OPONN equalization for various SNR levels. \textbf{b,} Experimentally obtained eye diagrams without (left) and with (right) OPONN equalization.}
    \label{fig:EQ}
\end{figure}

To demonstrate the versatility of our ultrafast nanophotonic processor, we next exploit it to compensate for nonlinear distortions that typically occur in a wireless communication channel. We assume that the input message to be transmitted contains a random stream of symbols $M(n)$ of the four-level-pulse-amplitude-modulation (PAM4) format. This modulation scheme is of practical interest in data centers and is implemented for instance in state-of-the-art 800G transceivers. Upon entering the communication channel, the message is subjected to multiple adverse effects that degrade the fidelity of the received signal $S(n)$ at the end of the channel. First, the presence of scatterers located around the path between the source and the receiver result in multipath fading that causes intersymbol interference among adjacent symbols. In addition, to overcome the losses that are incurred through the signal propagation, it is desirable to maximize the power of the transmitted signal by increasing the gain from the power amplifier in the output stage of the transmitter. This increase in the power is accompanied by nonlinear effects that distort the message. Additionally, the signal to noise ratio (SNR) is diminished by various noise sources, including thermal fluctuations and the noise introduced by amplifiers. Previous studies have shown the potential of recurrent neural networks in correcting for these cumulative distortions \cite{jaeger_harnessing_2004,vinckier_high-performance_2015,lupo_deep_2023}. Our objective is to perform this error correction task by exploiting the OPONN. Here, $N_\text{in}=5$ and $N_\text{out}=21$ nodes are used in the input and output layers, respectively. To assess the success of our approach, we calculate the symbol error rate (SER) defined as the percentage of the symbols detected incorrectly at the receiver end. Simulation results predict that the OPONN can effectively correct these errors and significantly improve the SER (Fig. \ref{fig:EQ}\textbf{a}). Our experiments corroborate this and show that a direct detection of the corrupted signal at the channel output results in a high SER of $19\% \pm 1.6\%$ (Fig. \ref{fig:EQ}\textbf{b}). In contrast, once trained, the OPONN is able to improve this result to $\text{SER} = 7\% \pm 2.3\%$. In order to confirm the role of the nonlinear processing performed by the OPO in error compensation, we also considered a purely linear equalization applied directly to the input pump, for which a $\text{SER} = 11\% \pm 1.8\%$ was measured. Given the fact that the measured SNR of the OPO output was lower than that associated with the input pump (please see Supplementary section 1), this latter comparison clearly indicates the efficacy of the OPO network to harness ultrafast photonic nonlinear processes to perform this equalization task. 

\begin{figure}
    \centering
    \includegraphics[width=1\textwidth]{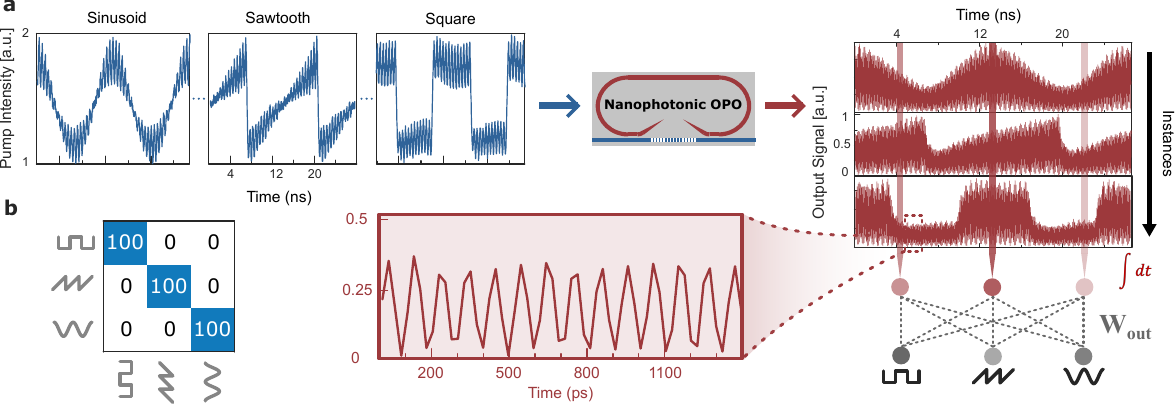}
    \caption{\textbf{Noisy waveform classification task. a,} We use the AWG to encode samples representing two cycles of periodic waveforms randomly chosen from three different classes of sawtooth, sinusoid and square signals with additive noise on the input pump (left). The measured signals at the output of the OPO associated with these waveforms are shown on the right. The output signals are sampled at three equally distant instances corresponding to three different nodes. The output layer of the OPO network is then set up by forming a fully-connected layer between these nodes and the output nodes that are activated based on a winner-takes-all approach. \textbf{b,} Using this simple architecture, we achieved 100\% accuracy for this classification task.}
    \label{fig:Waveform}
\end{figure}

One of the key applications in time domain signal processing involves the classification of various waveforms, with numerous applications ranging from bioinformatics \cite{min_deep_2017,dong_higher-dimensional_2023} to seismology \cite{mousavi_deep-learning_2022}. In order to showcase the generality of our approach, we next utilized the OPO neural network to distinguish among three classes of different waveforms randomly chosen from sinusoidal, square, and sawtooth signals that are contaminated with noise (Fig. \ref{fig:Waveform}\textbf{a}). Each waveform contains $N=50$ samples constituting two consecutive cycles. For this experiment, the OPONN is set up with $N_\text{in}=5$ nodes at the input. At the output of the chip, the  sub-harmonic signal generated by the OPO is sampled at three equidistant points, forming $K=3$ output neurons (Fig. \ref{fig:Waveform}\textbf{a}). The output layer of the neural network consists of a fully connected layer with $N_\text{out}=3$ nodes. During the training period, the linear weights associated with this output layer $\textbf{W}_\text{out}$ are optimized according to a winner-takes-all approach. We use a training set of 300 waveforms (100 from every class) and a different set of equal size for testing. In all cases, the OPONN was able to successfully classify all waveforms with $100\%$ accuracy, despite the presence of significant noise levels at the input (Fig. \ref{fig:Waveform}\textbf{b}).

\subsection*{Discussion}

In conclusion, we have demonstrated an ultrafast photonic neuromorphic processor based on a nanophotonic OPO chip operating at $10 \text{ GHz}$ clock rates. Our approach leverages coherent optical pulse propagation in conjunction with ultrafast parametric nonlinear processes available in TFLN to perform both the linear MAC operations and nonlinear activation functions in the optical domain. Operating at $\sim 10 \text{ GHz}$ clock rates, our OPO neural network is capable of achieving sub-nanosecond latencies, a timescale comparable to a single clock cycle in state-of-the-art digital electronic processors. We showcased the power of OPONN in performing a variety of machine learning tasks that involve time series signals including chaotic time series prediction, nonlinear channel equalization, as well as noisy waveform classification. In all cases, the OPONN achieved success rates exceeding 93\%. By eliminating successive slow and energy-intensive OEO conversions, our OPONN represents a significant leap towards realizing integrated all-optical neuromorphic processors operating at unprecedented high speeds while maintaining low energy consumptions.

\section*{Acknowledgments}
The authors acknowledge support from ARO grant no. W911NF-23-1-0048, DARPA award D23AP00158, NSF grant no. 1846273 and 1918549, Center for Sensing to Intelligence at Caltech, NASA/JPL, and NTT Research for financial and technical support. G.H.Y.L. acknowledges support from the Quad Fellowship.

\section*{Author Contributions}

All authors contributed to the writing of this manuscript.

\section*{Competing Interests}

The authors declare no competing interests with regards to the publication of this work.

\section*{Data Availability}

The data used to generate the plots and results in this paper is available from the corresponding author upon reasonable request.

\section*{Code Availability}

The code used to analyze the data and generate the plots for this paper is available from the corresponding author upon reasonable request.

\printbibliography

\end{document}